\documentclass[12pt]{article}
\usepackage{latexsym}
\usepackage[english]{babel}
\usepackage{fancyhdr}
\usepackage[mathscr]{eucal}
\usepackage[dvips]{graphicx}
\usepackage{graphics}
\usepackage{color}
\usepackage{epsfig}
\usepackage{amsmath}
\usepackage{amsthm}
\usepackage{amsfonts}
\usepackage{amssymb}
\usepackage{amscd}
\usepackage{bbm}
\usepackage{latexsym}
\usepackage{marvosym} 
\usepackage{pifont}
\usepackage{subfigure}
\usepackage{psfrag}
\usepackage{caption}

\begin{document}

\vskip 2 cm 

G.F.Dell'Antonio

Math. Dept. University Sapienza Roma 

and 

Math. Area, Sissa Trieste 

\bigskip

Contact interactions, Gamma convergence in Quantum Electrodynamics and the (Nelson) Polaron

\section{Introduction}

The  Polaron  is the quantum of a "Polarization" field [N] [L,S][M,V]. It is generated by the "contact" interaction [D1][D2] of a massive charged particle with  the  electromagnetic field. 

The  \emph{ contact  interaction} is a  process that takes place in a point. It is only defined in the "particles version" (photons and electrons) of Quantum Electrodynamics  and takes the place of "delta" interaction between Quantized Fields. 

In second quantization one may describe   polarization as follows.  A polarizable  massive charged particle (e.g. a spin $ \frac {1}{2} $ particle) emits a photon in a point. The interaction reverses the polarization of the photon and the spin of the particle. The photon is "immediately " reabsorbed.  

If the initial state of the field is not polarized the final result is the polarization of the massive  charged particle. The process can be described  as a \emph{contact interaction} of the charged particles with two photons, the one which is created and the one that is absorbed. The interaction is invariant under the Lorentz group. 

We prove that in this (particle) representation of the electromagnetic field (by photons)  the interaction produces, both in the relativistic and the non relativistic case, polarized bound states of the massive particle.   Each bound state is described by an  element of $L^(R^3$ times  the vacuum in the c.c.r. (canonical commutation relation)  representation of the photon  field which is \emph{inequivalent} to the Fock representation. 

In the relativistic case  representations associated to different  bound states are inequivalent.

Pictorially on can describe the bound states as a massive particle surrounded by a cloud made of an uncountable numbers of photons.

The bound states  are critical points of the energy functional (Pekar functional) [M,V]. Notice that in [M,V], as in our analysis, variational methods are employed and  perturbation theory  is not used. 

The process takes place in a point and can be repeated arbitrary many  times in different points.The result is the production of a \emph{Polarization Field}. 
The bound states are the \emph{Polarons} [N] [L,S][M,V].  

For the role played by  the representations of the canonical commutation relations (c.c.r.)  we call these bound states   \emph{Nelson Polarons} .

We study both the relativistic and the non relativistic case for the massive particle 
  
Since the number of massive particles is conserved, sectors with a different number of massive particles can be treated separately. We concentrate first on the one particle sector.

In the case of a single massive particle  we will prove that in the non relativistic case the Polarization Field is a field of single bound states with binding energy  proportional the strength $c$ of the interaction. In the relativistic case there are infinitely many  bound states with eigenvalues $ -c  \frac {1}{ \sqrt n }$ (so that the sum of the bound state energies diverges. 
 
 To each bound state corresponds a different representation of the canonical commutation relations (c.c.r.) of the photons; all are inequivalent to the Fock representation.

To describe mathematically the  "contact" interaction   we make  use as in [D1][D2]  of Gamma convergence [Dal] a variational tool introduced by E. de Giorgi in the sixties  in the study of finely fragmented materials.   This method  is only outlined here; it is presented  in more detains in [D1][D2].
 
It consists in  choosing as hamiltonian a suitable  self-adjoint extension of the free hamiltonian restricted to  function that vanish in a neighbourhood of the "contact manyfold". 

The method is non-perturbative and takes the place of renormalization  in the Fields presentation  of the electromagnetic field.

\bigskip

\section{Contact interaction and the existence of bound states} 

If the attractive interaction is sufficiently strong,  bound states may be present even if  the "region of interaction" is infinitely small (a point).

We shall consider zero range interactions, i.e contact interactions (we shall presently give precise definitions) .

These interactions produce bound states; they represent the \emph{polarizability} of the system and therefore it is natural  to call them \emph{Polarons}.

We shall describe them explicitly as non-Fock representation of the c.c.r. of the E-M field. For this explicit representation we call them \emph{Nelson Polarons} 

The spin structure of the massive particle is essential for the interaction to produce polarization; in fact this structure allows the interaction Hamiltonian to be a scalar.

The process is the following:  a massive charged quantum particles of spin $ \frac {1}{2} $ emits  a photon at a point (a contact interaction); the spin of the particle and the polarization of the photon are reversed (so that the interaction can be taken invariant under rotation) and then the photon is absorbed.  

This process can occur everywhere. The result is  a field of polariized particles, i.e. a field of Polarons.   
 
Since the interaction is by contact, we are  in the setting of contact interactions described in [D1][D2]. 

We will use the  \emph{contact interaction} instead of the delta function that describes interaction between field in  the Field Theory description of Electrodynamics.

And we will make use of Gamma convergence (a variational tool) instead of renormalization (a formal  procedure) 

 Let $ x_0, x_1, x_2 \in R^3 $ be the coordinates of the massive particle  and of the two mass zero particles  (photons)  with opposite orientations. 

Define  the symmetric operator $H_0$ as  the free hamiltonian restricted to functions that vanish in a neighbourhood of the \emph{contact manyfold}  $ \Gamma \equiv \{ x_0 = x_1 \} \cap \{ x_0 = x_2 \}$. 

Both particles have an "internal degree of freedom " (spin for the massive particle and polarizability   for the photon) but the hamiltonian is a scalar.  One has the vacuum as invariant state.

 For simplicity  in the following we neglect the "internal structures"  and we take all particles to be scalars. 
 
 Since the interaction takes place in a point, we are neglecting only a two-by-two matrix $ g \sigma_1$ where $g$ is strength of the interaction and the real Pauii matrix $ \sigma_1$ inverts the spin. 

In the non-relativistic the  free Hamiltonian of the zero mass particle is a second order differential operator

In the relativistic case it is  a differential operator of order one; in  the "scalar version" that we are using it is a  positive pseudo-diffential operator of order one. 

The  \emph{weak contact } hamiltonian  is \emph{defined}  as the self-adjoint extension extension that has in its domain  functions that take a fixed value $c$ at  $ \Gamma$. 

Integration by parts shows the the interaction potential is \emph{formally}  a (non tempered)  distribution supported by a point (in polar coordinates it is  a primitive of the delta function). It is not a bone-fide potential.

In the relativistic case the interaction is of \emph{strong contact}; in the domain of the extension there are function with a $ \frac {1}{ |x_i - x_j}$ behaviour at the boundary.

The potential is \emph{formally} a delta function. 

In both cases we must give an interpretation to the formal potential. 

Still  \emph{as quadratic form}  in both cases the potential is defined on continuous function and it is bounded  below.  

On the other hand  the free Hamiltonian in both cases is defines  on absolutely continuous functions a quadratic form which is positive (the difficulties came from integration by parts). 

Therefore in both cases the quadratic form of the  total Hamiltonian is bounded below and therefore \emph{if it defines a self-adjoint operator} this self-adjoint operator is bounded below.

It remains to be proved that the Hamiltonian exists and it is   a self-adjoint extension of the free Hamiltonian.  In both cases (weak and strong contact)  a constant $c$ (the strength of the interaction) defines the extension. 

The  interaction can take place in every  point and therefore these extensions are natural candidates to produce a field.

This field is a \emph{polarizabilty field}; we will see that it is a field  of bound states.

We shall prove that  in the non relativistic case (weak contact) there is only one  bound state with eigenvalue that is proportional to the coupling constant of the process. 

This bound state belongs to a non-Fock representation (which we give explicitly)  of the canonical commutation relations (c.c.r )  of the  free e.m.field. 

 in the relativistic case (strong contact) there are infinitely many bound states  with eigenvalues $- \frac {c}{\sqrt n}$ where $c$ isa function of  the coupling . 

The bound states can be expressed as product of a function $ \Phi_n  (x) \in L^2 ( R^3)$ times the vacuum of a representation of the zero mass field \emph{that depends on $n$ and on the strength  of the interaction} 

The field can be quantized; the quanta are the \emph{Polarons}. The name  \emph{Polaron}  is chosen since  they represent the "polarization  field"  due to creation and instantaneuos absorption  of a photon  at a  point.

We call them  "Nelson Polarons" because E.Nelson [N] was the first to describe the process in approximately  the same way as is presented  here.

\section{The field of Polarons}  

Each massive particle interacts separately with the mass zero field  and therefore we can restrict attention to the sector in which there is only one massive particle.

The entire system is translation invariant and we choose the reference  frame in which the interaction of the massive particle  takes pace at the origin. 

Later we comment briefly on the general case of $N$ massive particles. 

The expression we have given above for the interaction potential were  formal, since the "step function at $ Gamma$  and the "delta function" are not bona fide potentials. 

We shall give a precise meaning  to the interaction and discuss  in detail the self adjoint extensions.
 
We shall also prove  that the resulting Hamiltonian is  the limit, \emph{in strong resolvent sense} of Hamiltonians with potentials $ V^\epsilon  (x_i -x_j) , x \in R^3 $ that scale as $ V^\epsilon (|y|)  = c \frac { 1}{ \epsilon^2 } V ( \frac {|y|}{\epsilon}) $  in the non relativistic case and as  $ V^\epsilon (|y|)  = c \frac { 1}{ \epsilon^3 }  V ( \frac {|y|}{\epsilon}) $ in the  relativistic case. 

The Polaron is the "quantum" of  the "polarization field" due to the contact interaction.

All bound states can be represented as the massive particles  surrounded by a  "cloud" of uncountably many photons. 

Since photons are identical particles this cloud represents a \emph{Photon condensate}; the ground state is rotation invariant and  could be seen as  \emph{a ball  of light}.

We will prove that each bound state of the particle is associated to the vacuum of a  different (inequivalent)  representation of the canonical commutation relations (c.c.r) for the quantised zero mass field (recall that for zero mass field there are uncountably many  inequivalent representations.)

Different values of the coupling constant correspond to inequivalent representation. In the relativistic case the representations associated to different bound states  are \emph{inequivalent} and inequivalent to that of the non-relativistic case.  

 We treat separately the relativistic and non-relativistic cases. 

In absence of interactions there is no ground state (it would be the tensor product of  the vacuum of the field with a zero momentum  state of the particle) . 

If there is interaction the bound states are the product of a bound state of the massive particle times the vacuum of a suitable representation of the zero mass. The representation  depends on the bound state of the particle.  

We remark that in the non relativistic case only one bound state is produced.  In the relativistic case there infinitely many bound states with energies that scale as $ - c \frac {1}{\sqrt n}$. In both cases the the wave function of the bound states \emph{are not in}  $L^1 ( R^3)   $.

It is known that representations of the c.c.r. associated  with different states $ \Phi_1$ and $\Phi_2 $ are equivalent only is $ |\Phi_1 - \Phi_2|_1 < \infty $.  

In the present case  representations associated to different bound states \emph{are inequivalent}.

For each bound state $ \Phi (x) , x \in R^3 $  the representation  is \emph{the direct integral} of   representations  of the c.c.r., each associated to the value taken at $x$ by the wave function $ \Phi (x) $  of the bound state of the particle.

For the  proof we make use of the theory of self-adjoint extensions of symmetric operators and of Gamma convergence, a variational tool [Dal] [D2] introduced by E. de Giorgi in the study of finely fragmented composites.

As intermediate step in the proof,  in order to clarify the nature of the "contact potentials" , we use a map (we call it  "Krein map" ) that is "fractioning" and "mixing"; this is the reason why a natural tool is  Gamma convergence. 
 
\bigskip

\emph{REMARK } 

If there is no interaction there is no bound state. The energy of the bound state (or  of the lowest bound state in the relativistic case) can therefore be taken as (small) parameter $\epsilon $ that measures the strength of the interaction. It would is natural to take this strength as "perturbation parameter".

But the proof we give here is variational (through Gamma convergence of the resolvent of the Hamiltonian)  and not through  expansion of the Hamiltonian in series of powers of  this  small parameter.

We  prove that the hamiltonian of our system is the limit \emph{in strong resolvent sense} when $ \epsilon \to 0$ of hamiltonians in which the interaction is not a "contact " (zero range) interaction but is described by  potentials $ V^\epsilon (|x-x_k| ) = \frac {1}{ \epsilon^m} V( \frac { |x-x_k|} { \epsilon^m}) , \;\; k=1,2  \;\; x \in  R^3 $ where $ m = 2 $ in the non relativistic case and  $m = 3 $ in the relativistic case.

There is no convergence of quadratic forms and no convergence of a perturbation expansion in $ \epsilon $. Variational methods (Gamma convergence) are used to prove convergence of the hamiltonians.

  Notice that the photons have "mass zero"  and states with a non-denumerably   large number of photons  can still have finite energy and momentum. 

In the proofs we shall make use of compactness in the (Frechet`topology given by the  Sobolev norms \emph{and the number of massive of particles}. 
 
This allows the use  Gamma convergence  (a variational instrument originally introduced in Quantum Mechanics [D1][D2]) in place of \emph{renormalization}  (a formal scheme). 

For massive fields  as yet no topology has been found  that allows the use Gamma convergence as a substitute of  renormalization.  
  
Remark that in our system one of the particles is massive and  the analysis is done in each sector of $n$ massive particles. There is no natural \emph{field} associated to this massive particle. For  massive  particle there is no way to construct a  representation the c.c.r. that  is uniformly bounded  energy and momentum. 

It is plausible that the renormalization group (a scaling group of  the energies)  could be combined with Gamma converge to provide a non perturbative tool that can be used also in the massive case. It would perhaps correspond to the choice of a suitable representation of the c.c.r.  for each $N$ particle sector.   

 For completeness we mention that a Polaron  obtained by "abstract boundary conditions" is presented in [La].

 A precursor of the theory presented here may be found in the work of Dimock and Rajeev [D,Ra]. These Authors introduction  the "heat kernel  renormalization",  an operation that  for the potential coincides with   the Krein map we introduce here. But nothing is done to the kinetic part and (therefore)  Gamma convergence is not introduced.

\bigskip

...............................

\bigskip

\section{The non relativistic case} 

We  prove now that in the non relativistic case the weak contact (Nelson)  Polaron is a bound state and we give its wave function  $\Phi (x) $.
We prove also that $ \Phi (x) \notin L^1(R^3)$. 

We begin describing  weak  contact of the massive particle with any two of the zero mass non-relativistic particles.

The free Hamiltonian is  written  in Fourier transform $H_0 =  \frac {p^2} {2m} + p_1^2 + p_2^2  $.

Let $ H_0^0 $ be the symmetric operator defined by restricting  $H_0$ to functions that have support away from the "boundary" $ x= x_1 \cup x=x_2 $  

By definition a  hamiltonian of weak contact interaction is a self-adjoint extension of the symmetric operator $ H_0 ^0 $ (the operator $H_0$ restricted to functions that vanish in some neighbourhood  of the boundary). 

Weak contact extensions are parametrised by the value $c$ taken at the boundary by  functions in the domain. 
 
\emph{Formally}  they can be described by a step potential $ W_c$ that acts only at the boundary. 

The one-particle hamiltonian is  the "point interaction hamiltonian"" introduced in [A]. 
It is the limit as $ \epsilon \to  0$ of potentials $ V^\epsilon (x-x_1) $ that scale as $ V^\epsilon ( y)  = \frac {1}{ \epsilon^2} V ( \frac {y}{ \epsilon^2} ) ,  \;\, y \in R^3 $ with $ |V|_1 = C$

In $R^3$ the behaviour of the wave function at the boundary and the fact that the "potential" acts only at the boundary implies that there is a  zero energy resonance.

Recall that we are studying a system of a particle in interaction of two identical Bose particles.  This system  has no zero energy resonances since the symmetric product of the wave functions  of two zero energy resonances  is a bound state in the relative coordinates. Indeed when considering the inverse of the resolvent a symmetric two-by-two matrix with zero on the diagonal has one  negative (and one  positive) eigenvalue.  

Therefore the resolvent is regular at the origin and the extension has a bound state (and only one in the non relativistic case) . 

We study the system by using first, as intermediate step, a map (called in [D1[[D2] "Krein map" $ {\cal K}$ ) to a space of more singular function (called there "Minlos space" $ {\cal M}$). The map acts \emph{differently}  on the kinetic part (a self-adjoint operator) and on the potential term  (a quadratic form) and it is "mixing" ( with respect to the two particle channels).

For the kinetic part one has  $H_0 \to \sqrt {H_0} $. On the singular "potential" $ W_C $ at the boundary the map  it acts as $ W_C  \to W_C H_0^{-1} $.

One verifies that in $ { \cal M}$ the kinetic and potential terms have  the same "singular" action on functions that have support in a very small neighbourhood of the "contact manyfold  $ \{ x - x_1\} \cap  \{ x- x_2 \} = 0$.

It follows [D,R] that in $ {\cal M}$ there is a one-parameter family of self-adjoint operators each with a single  bound state; the bound states of the family cover the interval $ [-C, 0] $ 
 
Notice that in $ {\cal M}$  both the kinetic and the potential are represented by a family of self-adjoint operators.

Inverting  now the Krein map  one has in physical a  well ordered family of quadratic form bounded below.  

Due to the change in metric topology the forms are only weakly closed; they  are strictly  convex since the interaction affects only the s-wave.

We have noticed that the "Krein map" is "fractioning" (the space $ {\cal M}$ is a space of more singular functions)  and  "mixing" (it is not diagonal in the particle "channels").

This suggest to make use of Gamma convergence [Dal], a variational method introduced by E. de Giorgi  in  the study of homogenisation i.e. the study of finely fragmented materials (for a variational analysis of the Polaron problem see [M,V]).

In a topological space $Y$ the Gamma limit of a sequence of strictly decreasing convex quadratic  forms $ F_n $ is the quadratic form $ F(y)$  characterised by the following relations

\begin{equation} 
\forall y \in  Y, y_n \to  y, F(y) = lim inf F_n(y) \;,\; \forall x \in  Y _n \forall  \{ x_n  \} : F(x) \leq  limsup_{n} F_n(x_n) 
\end{equation}

 The first condition implies that $F$ is a common lower bound for the  $F_n$ , the second implies that the bound is optimal. 

In our case the Gamma limit exists because there are no zero energy resonances and therefore the quadratic forms belong to  a compact set for the Frechet topology given by the Sobolev semi-norms.

Gamma convergence selects the most negative form (to which by definition the others \emph{Gamma converge } ); being a minimum this limit form can be closed strongly [K] and defines a self-adjoint operator $ \hat H_c $ (the Gamma limit). 

Recall that  $C$ is the value taken at the boundary by functions in the domain of the chosen extension.  

A strictly decreasing sequence in a compact region of a Frechet space has a unique limit.

Gamma convergence implies resolvent convergence [Dal]  \emph {but not convergence of the quadratic forms}; the  sequence of quadratic forms converges only if it  is uniformly bounded.

It it easy to verify that $ \hat H_c $ is also the Gamma limit of the approximating Hamiltonians 

\begin{equation} 
H^\epsilon = H_0 + \int V^\epsilon (x-y_1)  \Psi (y_1)  dy_1 + \int V^\epsilon (x- y_2)  \Psi (y_2)  dy_2 
\end{equation} 

with potentials that scale as $ V^\epsilon (x-y) = \frac {1}{\epsilon^2} V ( \frac {x-y} {\epsilon}) $ of constant $L^1$  norm  $ |V(x)|_1 = C $. 

Indeed the sequence of $\epsilon$-dependent quadratic forms is  a decreasing sequence that has as lower bound the quadratic form $ H_C$.
Therefore $ H_c$  is the limit, in strong resolvent sense, of the sequence of hamiltonians $ H_\epsilon$.

  Denote by  $ \Psi _\epsilon  (x), x \in R^3  \times R^3) $ the bound state for the approximation hamiltonian. 

One has  $ |\Psi_\epsilon - \Psi |_2 \to 0$ in $L^2 (R^3 \times R^3 ) $ but  the limit  vector is the symmetric part of the product of two zero energy resonances and therefore it is not in $ L^1( R^3 \times  R^3) $.

Therefore for any value of the parameter $c$(the value at the boundary)  the representation we have obtained for  the field of zero mass particles is not equivalent to the Fock representation and to different values of $c$ correspond \emph{inequivalent} representations of the c.c..r.  (recall that it is the $L^1$ norm and not the $L^2$ norm that provides equivalence of representations).

\section{The relativistic case}

The difference with the previous case is that  here also the hamiltonian of the relativistic particles is a first order differential operator. 

The free hamiltonian is now $ \sqrt {|p|^2 + m} +  |p_1| + |p_2|$ .

The massive particles  is in weak contact separately with the other two. 

The  weak contact is still defined by the requirement that the function of the domain of the extension take a finite value at the boundary. 

This is the same condition as in the non-relativistic case but now the kinetic energy is a first order (pseudo-differential ) operator.

Also here we can take  as reference the frame in which the particle of mass $m$ is at rest. 

We perform  next  a separate \emph{change of coordinates} for the relativistic particles. In the description  in polar coordinates   $ \{ p, \theta, \phi \} , \; p \in R^+$ we take as new coordinates  $ P = \sqrt p  $ so that $ p= P^2 $.

This turns the relativistic two-particle hamiltonian into the non-relativistic form;  it turns  weak contact into strong contact.  Indeed changing the coordinates changes also the form  of the boundary condition. 

\emph{Formally} this interaction is described by a delta function. 

Once again we introduce a Krein map [D1] [D,2], a map  which is mixing and fractioning and acts differently of the kinetic and potential terms and does not break rotational invariance. We recall  briefly its structure.

The action  on the kinetic is $ H_0 \to \sqrt{H_0} $. The action on the potential  term is $ V \to V \frac {1}{ H_0 } $ where $ V = \delta (x-x_1) + \delta (x -x_2)$.

The "Minlos space" ${ \cal M}$  is a space of more singular functions.

In $ {\cal M}$ the kinetic term is $ \sqrt H_0$ and the potential term has in each channel a pole singularity. in position variables. 

Therefore [D,R] if the potential is strong enough there is a one parameter family of self-adjoint operators unbounded below each with an infinite number of bound states with eigenvalues that that scale as $ - \frac {1}{n} $. 
  
"Inverting" the Krein map one has now , due to the change in metric topology,  a one parameter family of \emph{weakly closed} well ordered strictly convex quadratic forms bounded below; the forms are strictly convex because the interaction modifies only the s-wave. 
 
 Again the condition for existence of the Gamma limit is that the sequence be contained in a compact set for the topology of $Y$ (so that a Palais-Smale converging sequence exists). 

Once again the topology is the  Frechet topology defined by the Sobolev semi-norms and compactness is assured also in this case by the absence of zero energy resonances. 
Therefore the Gamma limit exists. 

We use  Gamma convergence to select the infimum.
 
Since it corresponds to a minimum by a theorem of Kato [K] the limit form admits strong closure and  defines a self-adjoint hamiltonian.
It represents separate \emph{separate strong contact}   of a particle of mass $m$ with two non relativistic zero mass particles. 

It is a self-adjoint operator with an infinite number of bound states  $ \Phi_n (x )  \;\;\; , n =0,1,  ,  ...$ where $ x $  are the coordinate of the massive  particle.  

The energies of the bound states scale as $ -c \frac {1}{\sqrt n}$. Only the ground state has a wave function that can be chosen real. 

Notice that the choice of different coordinates for the relativistic particle  does not change the metric of the massive particle. 
 
 Each  bound state $ \Phi_n $ defines a representation of  the  field of the identical zero mass particles;  the representations are inequivalent since  $ \Phi_n  - \Phi_m \notin L^1 (R^3)$ for $ n \not= m$.        

The negative part of the spectrum of the system is now given by an infinite collection of states,  a state of the massive particle  with wave function $ \Phi_n  $ decorated with a cloud of zero mass particles. The representations associated to different bound state are inequivalent. 

Each representation  is the direct integral of irreducible representations of the zero mass field parametrised by $x \in R^3$ and associated to  the value  $ \Phi_n $  at the point $x$ of the wave function of the particle of mass $m$. 

\bigskip

\section{ Representations of the zero mass field associated to a bound state of the massive particles.}

Here we give the representation  of the zero mass  field (photons) associated to the bound states of the massive particles. 

In the reference frame  we have chosen the wave function of each bound state has its barycentre  at the origin.

 Let $ \Phi (x) $ be the wave function of the bound state normalised to  $ \| \Phi  \! _2 = 1 $

One can  associate to the wave function $ \Phi (x) $ a representation of the field  which is the   direct integral over $x$ of the representation  of the c.c.r. in which the field of annichilation operators $A_x^\Phi $ is defined by 

\begin{equation}
 A_x^\Phi   (y) = a (y)  -  \Phi(x ) 
 \end{equation} 

where the field $a(y) $ is in the Fock representation.

In these notation the field is $ \Phi (x) = A(x) + A^* (x) ,  \;\;  A(x) = A_x^\Phi$

 For each value of $x$ the  operators $A_x^\Phi (y)$ and  $(A_x^P(y) )^*$ satisfy the same c.c.r. as the operators $a(y), \; a^* (y) $ but  it is known that the  representations associated to two different wave functions $ \Phi_1 ,\;\;   \Phi_2 $ are \emph{inequivalent} if $ \Phi_1 -\Phi_2 \notin L^1$. 
 
Since the coupling with the field in linear if one writes the Hamiltonian as a function of the field $A_x^\Phi   (y) $ one obtains 

\begin{equation}
H  = H_0   + \int \omega (p)   (A_x^\Phi) ^* (p) A_x^\Phi  (p)dp 
\end{equation}

(in the Theoretical Physics  Literature this operation takes the name of "completing the square".

In order to minimise the energy one must choose for every $x$ the vacuum of the  Fock representation  for $A_x^\Phi  $ 

The ground state $ \Omega (x) $ satisfies then  for every value of $x$ the equation $ A_x^\Phi  \Omega  = 0  $

In the reference frame we have chosen the  self-adjoint operator $\hat H $ has as ground state the massive particle with wave function $ \Phi (x)$ times the direct integral over  $x$ of representations that    \emph{depends on the coordinate of the particle of mass one}. [N] [G,W], [L,S] , [S].

\section{ The case of dimension two } 

We have treated the case of dimension 3 ; a similar analysis of the Polaron can be made for dimension two; here the delta function has the same scaling properties as the laplacian and therefore in this sense the contact is weak. There are no zero energy resonances. 

Separate contact interaction with two particles can be analysed as in three dimensions. 

The Krein map is again constructed with $ \sqrt H_0$ and is again mixing and fractioning.

 Gamma convergence  is used as in the case of strong contact in three dimensions.

In two dimensions the Polarization field is a field of single bound states.

\bigskip

\section{On the particle formulation  (photons and electrons) of Quantum Electrodynamics}

In the previous analysis we have introduced photons (zero mass particles)  i.e. we have  adopted the particles formulation of Q.F.T.

As a consequence we have introduced interactions that in same sense "substitute" interactions between fields, i.e "contact interactions" between particles.

Contact interactions are "local interactions" that take the place  of a "delta potential"  (that is ill defined for particles). 

In a description in which particles are represented  by functions (probability waves)  local interactions are  self-adjoint extensions of the  symmetric operator defined as the free hamiltonian restricted to functions that vanish in some neighbourhood  of a "contact manyfold ".

The analysis we have made here extends trivially to the separate contact interaction of the field of  zero mass particles  with  an arbitrary (but finite) number N particles of mass $m > 0  $ i.e to Quantum Electrodynamics in its particles version.  

Since the interaction takes place in a point and the particles are represented by functions in $L^2(R^3)$, we may safely assume that different massive particles interact 
separately and independently with two distinct pairs of zero mass particles, the photons. 

The massive particles are can be relativistic or non relativistic. Correspondingly the  contact interaction with the photon may be weak or strong

In the particle interpretation of the zero mass field there may be states of bounded energy and momentum that contain uncountably many particles.

Traditionally one calls Fock representation (or Fock sector) the one  in which there is a countable number of zero mass particles. The states of this representation are obtained by acting countably many times on the vacuum with an operator that adds a (zero mass) particle.  

We have proved that there are  states in which the massive particle is surrounded by a cloud of zero mass particles. In the non relativistic case there is only one such state, 
in the relativistic case there is a infinite number. These clouds are not in the Fock sector and are therefore they cannot be found in .Fock space analysis of the field. 

A trace of them  remains in the "polarizabilty field" and in its quanta, the "Polarons"

We stress that these  results are obtained  in the "particle"  formulation of Quantum Electrodynamics  in which the interaction is not described  by a delta function but rather a contact interaction, weak or strong

We have considered only the one-particle sector. The analysis extends to sectors of an arbitrary number of particles because sectors corresponding to different number of massive particles are not coupled by the contact  interaction.

The Hamiltonian is the limit of hamiltonians with potentials with support that shrinks to a point, but no rate of convergence can be given in the parameter that controls the size of the support.

One can add to the contact interaction any regular interaction  [K.K] without altering the results obtained with contact interactions. 

 We stress that the method of analysis we follow here is \emph{variational and not perturbative}. 
 
 Gamma convergence is  a substitute for \emph{renormalization theory} , a formal procedure that is used in  the \emph{Field Theory version} of electrodynamics.

 We recall that it is argued  in [D1[[D2] that the semiclassical limit of weak contact interaction is Coulomb interaction and that this is also the non relativistic limit  

Subtraction of a zero point energy is the only renormalization needed in the Field Theory formulation of Quantum Electrodynamics in the non relativistic case.

\section{References}

A]  S.Albeverio,R. Hoegh-Krohn,Point Interaction as limits of short range interactions on Quantum Mechanics   J. Operator Theory   6 (1981) 313-339 

[D1]  G.F.Dell'Antonio   Contribution to the volume in honour of S.Albeverio Springer 2020  
 
[D2]  GF.Dell'Antonio  Contact interactions in Q.M., Gamma convergence and Bose-Einstein condensation  arXiv.2003.10755

[Dal]  G.Dal Maso Introduction to Gamma-convergence Progr Non Lin. Diff Eq. 8, Birkhauser (1993)  

[D,R] S.Derezinky, S.Richard  On the Schr\"odinger operator with inverse square potential on the positive real line. Arkiv  1604 03340

[D,Ra] J.Dimok,S.Rajeev  Point interaction on three and two dimensional manifolds   J. Physics A 37 (2004) 9157b

[L,S]  E.H.Lieb ,L.Seiringer Equivalce of two definitions of the effective mass of the Polarin. J.Stat.Phys. 154 (2014) 51

[La] J.Lampart, On a direct description of the pseudo-relativistic Nelson polaron  arXiv:1810.03313

[M,V]  C.MukherYee , S.R.S. Varadhan Identification of the Polaron measure in strong coupling and the Pekar variational formula arXiv : 1812.o6927 

[N] E.Nelson    Interaction of non relativistic  particle with a quantized scalar field. J.Mat.Phys. 5 (1964)b1190

\end{document}